# Effective Hamiltonian of strained graphene

T L Linnik

Department of Theoretical Physics, V E Lashkaryov Institute of Semiconductor Physics, Kyiv 03028, Ukraine

**Abstract.**

Based on the symmetry properties of graphene lattice, we derive the effective Hamiltonian of graphene under spatially non-uniform acoustic and optical strains. We show that with the proper selection of the parameters, the obtained Hamiltonian reproduces the results of first-principles spectrum calculations for acoustic strain up to 10%. The results are generalized for the case of graphene with broken plane reflection symmetry, which corresponds, for example, to the case of graphene placed at a substrate. Here, essential modifications to the Hamiltonian give rise, in particular, to the gap opening in the spectrum in the presence of the out of plane component of optical strain, which is shown to be due to the lifting of the sublattice symmetry. The developed effective Hamiltonian can be used as a convenient tool for analysis of a variety of strain-related effects, including electron-phonon interaction or pseudo-magnetic fields induced by the non-uniform strain.

1. **Introduction**

Unique properties of graphene combined with a break-through in technology of its preparation inspired very extensive research activity in the field [1-5]. It is well understood that effect of strain on the graphene properties is very important. First of all, most often graphene is deposited or grown on some substrate [6-12]. In this case strain appears naturally due to coupling of graphene with the substrate. Apparently, this static strain perturbs the spectrum of graphene, which have strong effect on its properties. Then, dynamic strain due to the graphene lattice vibrations is a basis of electron-phonon interaction which controls the electron transport characteristics of graphene. Recently especial attention is paid to the non-uniformly strained graphene where such strain is equivalent to creation of the pseudo-magnetic fields [13-17].

As in bulk crystals, two distinct approaches can be applied to study the effect of strain in graphene. The first one is based on some microscopic model in the frames of which the carrier spectrum of the strained graphene is calculated. In particular, the tight-binding model or first principles calculations are used for this purpose [18-25]. Alternatively, using many-band kp method it is possible to develop strain-dependent Hamiltonian of graphene. Restricting ourselves by the terms of particular power of strain magnitude and using the symmetry considerations we can derive possible form of Hamiltonian. This approach can be applied efficiently for smooth, on the scale of the lattice constant, strains. Usually, two kinds of strain are distinguished: "acoustic" and "optical" [26]. In the former case both carbon atoms forming the elementary cell are displaced in the same way, while the latter case corresponds to the relative shift of such atoms, which leaves the position of the center of mass unchanged. Of course, such Hamiltonian contains unknown constants, which should be obtained from experiments or first-principles microscopic calculations. The advantage of this method is its applicability to a broad spectrum of strain-related effects, including modification of spectrum, electron-phonon interaction, etc. Many-decade studies of bulk crystals proved its efficiency. For unstrained graphene, such a symmetry based approach was developed in [27] and [28].

For most dielectrics and semiconductors considering the deformation potential Hamiltonian it is possible to restrict ourselves by the terms which depend on strain only [29]. In graphene the terms dependent on both strain and wavevector should be retained since they lead to essential modifications of strain effects. This fact is sometimes overlooked in the literature. The extensive study of possible effective Hamiltonians in graphene was done in [30], including acoustic strain $k$-dependent terms. In this work we, with the use of kp-expansion, analyze the origin of these terms, which allows us to estimate the values of the parameters of Hamiltonian. We show that these estimates agree well with the quantitative analysis made by comparison with the results of first principles calculations [21]. Then, we derived the effective Hamiltonian for the case of optical strain, where, to the best of our knowledge, only k-independent contributions were analyzed so far [13]. In addition, we address the case of reduced graphene symmetry, namely where there is no reflection symmetry in the plane of graphene sheet. Physically, this corresponds to graphene coupled to a substrate by van der Waals forces, or to suspended graphene in normal electric field. In unstrained case, Dirac spectrum survives under this reduction of symmetry. However, we show that absence of this reflection symmetry has essential implications for the strain effects. In particular, a gap opens in the spectrum under the out of plane optical strain. We show that appearance of the gap is due to breaking of the sublattice symmetry under such strain.

The paper is organized as follows. In Section 1 the standard approach introduced by Bir and Pikus for strain-induced terms in the kp-Hamiltonian is applied to the graphene layer and the corresponding terms are derived with the use of the invariant method. Section 2 is aimed to analyze the spectrum of strained graphene and, comparing it to the results of first-principles calculations, to extract the constants determining the acoustic-strain related terms in the Hamiltonian. In Section 3 we obtain the effective strain Hamiltonian for the case of reduced symmetry and analyze its implications. Section 4 is devoted to generalization of our results for the case of unhomogeneous strain.

.

2. **Hamiltonian of graphene under the homogeneous strain**

In this paper we consider two types of strain referred to as acoustic and optical strain. Such deformations appear in long-wavelength acoustic and optical phonons, respectively [26,29]. In the former case the two atoms of graphene unit cell shift identically and strain is described by the strain tensor $\varepsilon_{ij}$. The latter kind of deformation is characterized by no displacement of the unit cell center of mass. Since in graphene the two atoms of the unit cell are identical, it is characterized by their displacements, $\mathbf{u}_{1,2}$, as $\mathbf{u} = \mathbf{u}_2 - \mathbf{u}_1$. In general case the strain destroys the Bravais lattice of the crystal and usual perturbation theory can not be applied. To avoid this complication one introduces the deformed coordinates which rebuilt the Bravais lattice to the unstrained one [29,31-34]. As a result of such transformations, the Hamiltonian includes strain-induced terms, originating from perturbation of both kinetic and potential energy. For smooth strain, the standard effective mass scheme can be applied. As a result, the effective Hamiltonian has the following form [29,31-34]:

$$H = H_k + H_\varepsilon + H_{\varepsilon k} + H_u + H_{uk},$$

$$(H_k)_{n'n} = \frac{\hbar}{m_e}(\mathbf{p}_{n'n} \cdot \hat{\mathbf{k}}), \qquad (H_\varepsilon)_{n'n} = \sum_{ij} D^{ij}_{n'n}\varepsilon_{ij},$$

$$(H_{\varepsilon k})_{n'n} = -\frac{\hbar}{m_e}(\mathbf{p}_{n'n} \cdot \delta\hat{\mathbf{k}}) + \frac{\hbar}{m_e}\sum_s \frac{(\hat{\mathbf{k}}\mathbf{p}_{n's})(H_\varepsilon)_{sn} + (H_\varepsilon)_{n's}(\hat{\mathbf{k}}\mathbf{p}_{sn})}{E_n - E_s}, \qquad (1)$$

$$(H_u)_{n'n} = -(\mathbf{u} \cdot \nabla_\mathbf{u} V_{n'n}),$$

$$(H_{uk})_{n'n} = -\frac{\hbar}{m_e}\sum_s \frac{(\hat{\mathbf{k}}\mathbf{p}_{n's})(\mathbf{u} \cdot \nabla_\mathbf{u} V_{sn}) + (\mathbf{u} \cdot \nabla_\mathbf{u} V_{n's})(\hat{\mathbf{k}}\mathbf{p}_{sn})}{E_n - E_s}.$$

Here $n$ and $n'$ mark the electron and hole bands of graphene, and summation is over remote bands, $s$. $H_k$ is a standard kp Hamiltonian of unstrained crystal characterized by the lattice potential $V(\mathbf{r})$, which results in the cone of electron and hole bands at points $K$ and $K'$ of the Brillouin zone of graphene. The rest of notations used in (1) are $\hat{\mathbf{k}} = -i\nabla$; $\delta k_i = \sum_j \varepsilon_{ij} k_j$ which can be interpreted as the acoustic strain-induced change of wave vector, and $D^{ij}_{m'm}$ are the deformation potential constants, determined by the following expression [29,31-34]:

$$D^{ij}_{m'm} = -\frac{(p_i p_j)_{m'm}}{m_e} + \left.\frac{\partial V}{\partial \varepsilon_{ij}}\right|_{\varepsilon_{ij}=0} \qquad (2)$$

The two terms in the expression for $H_{\varepsilon k}$ correspond to the first and second order perturbation theory which appear due to perturbation of the kinetic and potential energy, respectively [29,34]. For $H_{uk}$, only second-order terms are present. The survival of the first term in $H_{\varepsilon k}$ is related to Dirac spectrum of graphene. In fact, it is determined by the same matrix element $p_{n'n}$, as $H_k$, which allows us to quantify the corresponding contribution to $H_{\varepsilon k}$. Then, if we suppose that the interband deformation potentials are of the same order as the intraband ones, which are determined to be about 20 eV [35,36], and taking into account that the energy distance to the nearest band at K point is about ~10 eV [37], we find that the contributions of the second-order perturbation terms in $H_{\varepsilon k}$ can be comparable with that of the first-order.

Typically, $H_{\varepsilon k}$ and $H_{uk}$ are not considered while treating strain effects [29]. This is because their contribution is less than that of $H_{\varepsilon}$ and $H_{u}$. However, as we will show below, $H_{\varepsilon k}$ and $H_{uk}$ give rise to qualitatively distinct properties of graphene spectrum which do not appear due to $H_{\varepsilon}$ and $H_{u}$. Note that similar situation arises in semiconductors, where $H_{\varepsilon k^2}$ contribution can be manifested in narrow quantum wells and wires [38].

As it is known, the general form of effective Hamiltonian (1) can be written solely on the basis of symmetry considerations using the invariant method [29,30,34]. Within this method one finds possible Hamiltonian terms which are invariant with respect to the symmetry transformations of the crystal lattice, which should be, in addition, tested for the time-reversal invariance. The honeycomb lattice of graphene is illustrated in Fig.1(a). The corresponding first Brillouin zone is a hexagon and zone extremum is realized at the two inequivalent corners of the hexagon K and K' (Fig. 1(b)). The wave vector point group in the K (or K') point is D$_{3h}$ and includes the rotation by $2\pi/3$ around the $z$-axes, which is normal to the graphene sheet, reflection $\sigma_v$ in the plane perpendicular to the $y$ axes, and reflection $\sigma_h$ in the plane of the graphene sheet. Note, that $\sigma_v$ reflection is present due to equivalence of the two sublattices composing graphene lattice (see Fig. 1(a)). Applying the method of invariants, we can rewrite Hamiltonian (1) in the vicinity of K-point as:

$$
\begin{aligned}
H_k &= \hbar v_F ((\hat{k}_x - i\hat{k}_y)\sigma_+ + (\hat{k}_x + i\hat{k}_y)\sigma_-), \\
H_\varepsilon &= E_{d1}\bar{\varepsilon} \cdot I + E_{d2}((\varepsilon_\Delta + 2i\varepsilon_{xy})\sigma_+ + (\varepsilon_\Delta - 2i\varepsilon_{xy})\sigma_-), \\
H_u &= E_{u1}((u_y + iu_x)\sigma_+ + (u_y - iu_x)\sigma_-), \\
H_{\varepsilon k} &= \hbar v_F [a_2((\hat{k}_x + i\hat{k}_y)(\varepsilon_\Delta + 2i\varepsilon_{xy}) + (\hat{k}_x - i\hat{k}_y)(\varepsilon_\Delta - 2i\varepsilon_{xy}))I + \\
&\quad + (2d_2 - 1)\bar{\varepsilon}((\hat{k}_x - i\hat{k}_y)\sigma_+ + (\hat{k}_x + i\hat{k}_y)\sigma_-)/2 + \\
&\quad + (2g_2 - 1)((\hat{k}_x + i\hat{k}_y)(\varepsilon_\Delta - 2i\varepsilon_{xy})\sigma_+ + (\hat{k}_x - i\hat{k}_y)(\varepsilon_\Delta + 2i\varepsilon_{xy})\sigma_-)/2], \\
H_{uk} &= E_{u2}((u_y - iu_x)(k_x + ik_y)\sigma_+ + (u_y + iu_x)(k_x - ik_y)\sigma_-).
\end{aligned}
\qquad (3)
$$

where $v_F = 10^6 m/s$ is Fermi velocity which determines the spectrum of the unstrained graphene and $a_2$, $d_2$, and $g_2$ correspond to contributions of second-order kp-perturbation theory, $\sigma_\pm = (\sigma_x \pm i\sigma_y)/2$ are the combinations of Pauli matrices, $I$ is $2\times 2$ unity matrix, $k_\pm = k_x \pm ik_y$, and we introduce the uniaxial and hydrostatic components of acoustic strain: $\varepsilon_\Delta = \varepsilon_{xx} - \varepsilon_{yy}$, $\bar{\varepsilon} = \varepsilon_{xx} + \varepsilon_{yy}$. Some terms of Hamiltonian (3) were identified previously. So, the term $H_\varepsilon$ was considered in [30, 35] and [36] and the deformation potentials were deduced from experimental data and first-principles calculations [35,36]. $H_u$ term was addressed in [13]. In a recent paper of Winkler [30] $H_{\varepsilon k}$ invariants were introduced. Our analysis from Eq.(1) allows us to trace the origin of different terms entering into $H_{\varepsilon k}$. The invariants originating from the first term due to kinetic energy perturbation are determined by the same momentum matrix element as $H_k$, and it can be obtained from the latter by the substitution $\mathbf{k} \to -\delta\mathbf{k}$. However, the same invariants appear due to the second term. As a result, two last terms in $H_{\varepsilon k}$ in Eq. (3) contain contributions from both first, $\hbar v_F$, and second-order, $\hbar v_F g_2$ and $\hbar v_F d_2$, kp perturbation theory. Below, using the results of the first-principles calculations [21] we will show that, in accordance with our previous estimate, the magnitude of $g_2$ and $d_2$ is of the order of unity. The term $H_{uk}$ was not considered so far. It is worth to mention, that under the reflection $\sigma_v$, which change the positions of the two atoms of the unit cell, the displacement $\mathbf{u}$ has additional change of its sign, compare to the case of a usual vector [26,29]. As a result, the corresponding invariant differs from the quadratic vector invariant given in [30].

The provided consideration is valid, strictly speaking, to a suspended sheet of graphene. In most cases, however, graphene is placed at some substrate [6-12]. It is supposed usually that the mechanically exfoliated graphene is bounded to substrate by van der Waals forces [9,10,39,40]. With a good accuracy, this coupling is laterally isotropic. From the symmetry point of view, this means absence of the reflection $\sigma_h$, while the rest of the symmetry elements of the wave vector point group are preserved. Similar situation arises if suspended graphene is placed in external normal electric field. As a result, the wave vector point group lowers to $C_{3v}$. It is important, that for this truncated symmetry $H_k$ has the same form conserving Dirac spectrum. We have found that $H_\varepsilon$, $H_{\varepsilon k}$, and $H_{uk}$ also has the same form as in Eq.(3), and an additional contribution $H_u^{(s)}$ appears:

$$H_u^{(s)} = E_{op} u_z \sigma_z. \tag{4}$$

As we see, the new term contains out of plane component of optical strain. Note that considering $H_\varepsilon$ we omitted the term proportional to $\varepsilon_{zz}$ since this component of strain tensor does not have physical meaning for single layer graphene.

It is worth to mention that Hamiltonian (4) is in accordance with more general symmetry considerations. Indeed, $z$-component of optical strain in graphene at a substrate lowers symmetry of graphene to that of a honeycomb lattice with nonequivalent sublattices and no $\sigma_h$ reflection symmetry. In this case the wave vector point group at point K (or K') of Brillouin zone is $C_3$ and it has only nondegenerate representations. Thus, the energy states of such a system are nondegenerate. It is straightforward to obtain the two-band Hamiltonian, considering the two nearest bands, which after transferring the symmetry to $C_{3v}$ merge to the Dirac cone. It means that the perturbation breaking the sublattice symmetry is supposed to be weak. Again, with the method of invariants, after proper rotation of $k$ frame we find it to be

$$H_h = \alpha((\hat{k}_x - i\hat{k}_y)\sigma_+ + (\hat{k}_x + i\hat{k}_y)\sigma_-) + \beta\sigma_z, \tag{5}$$

where $\alpha$ and $\beta$ are some real constants. One can see that $H_h$ is equivalent to $H_k + H_u^{(s)}$. Note, that the form of Hamiltonian (5) does not change in the presence of $\sigma_h$ symmetry. Hamiltonian (5) is the particular case of the Haldane model [41] where two sublattices of honeycomb lattice are assumed to be inequivalent and which can be realized, for example, in graphene placed at the BN substrate [42]. As we see, the normal optical strain on graphene at a substrate can serve as another feasible realization of such a model.

Another important modification is that in this case out of plane component of acoustic *displacement* $\bar{\mathbf{u}} = (\mathbf{u}_1 + \mathbf{u}_2)/2$ also appears in the Hamiltonian. In bulk crystals there is no such contribution since displacement of the whole crystal cannot modify the band spectrum [43]. This applies also for the case of suspended graphene. However, under no $\sigma_h$ symmetry this argument does not work for out of plane displacements. Considering possible invariants, we obtain the corresponding terms of Hamiltonian:

$$H_{\bar{u}}^{(s)} = E_{s1}\bar{u}_z I, \qquad H_{\bar{u}k}^{(s)} = E_{s2}\bar{u}_z((k_x - ik_y)\sigma_+ + (k_x + ik_y)\sigma_-). \tag{6}$$

As we will show in the following section, some out of plane contributions give rise to important modification of the graphene spectrum. It is worth to mention that since the surface of graphene is corrugated [44-47] considered out of plane contributions can be essential. For the case of suspended graphene in external normal electric field the constants in Eqs. (4) and (6) are proportional to the magnitude of the field.

### 3. Spectrum of graphene under the acoustic and optical strain

Recently there were a great number of publications where the effect of uniform strain on energy spectrum of suspended graphene sheet was explored within a tight-binding approach and first principles calculations [18-25]. The main results are that the opening of a gap in the energy spectrum requires very high values of deformation of the order of 20%. This means that the energy spectrum remains gapless and cone-like for small strain. However, the Dirac points in strained graphene no longer coincide with the edges of the Brillouin zone, $K$ and $K'$. Besides, the strong Fermi velocity anisotropy was observed [18,20,21]. All these results can be easily reproduced with the use of the obtained effective Hamiltonian, Eq.(3). It is straightforward to obtain the energy spectrum:

$$E_k = \Delta E + 2\hbar v_F a_2[(k_x - k_{x0})\varepsilon_\Delta - 2(k_y - k_{y0})\varepsilon_{xy}] \pm$$
$$\pm \sqrt{\alpha_x(k_x - k_{x0})^2 + \alpha_y(k_y - k_{y0})^2 + 2\alpha_{xy}(k_x - k_{x0})(k_y - k_{y0})}. \tag{7}$$

Here

$$k_{x0} = -(E_{d2}\varepsilon_\Delta + E_{u1}u_y)/\hbar v_F, \quad k_{y0} = (2E_{d2}\varepsilon_{xy} + E_{u1}u_x)/\hbar v_F,$$

$$\Delta E = E_{d1}\bar{\varepsilon},$$
$$\alpha_x = (\hbar v_F)^2(1 + \tilde{d}_2\bar{\varepsilon} + \tilde{g}_2\varepsilon_\Delta) + 2\hbar v_F E_{u2}u_y,$$
$$\alpha_y = (\hbar v_F)^2(1 + \tilde{d}_2\bar{\varepsilon} - \tilde{g}_2\varepsilon_\Delta) - 2\hbar v_F E_{u2}u_y, \tag{8}$$
$$\alpha_{xy} = 2\tilde{g}_2(\hbar v_F)^2\varepsilon_{xy} + 2\hbar v_F E_{u2}u_x,$$

where notations $\tilde{g}_2 = 2g_2 - 1$ and $\tilde{d}_2 = 2d_2 - 1$ are used and only linear in strain contributions to coefficients $\alpha_i$ are considered. As we see from the Eq. (7), in this approximation the spectrum is gapless. To open the energy gap the higher-order quadratic in strain terms should be taken into account in Eq. (3). This is in line with the results of first principles and tight-binding calculations suggesting the high threshold value of strain for gap opening.

It is convenient to rewrite Eq. (7) in the form showing explicitly the angular anisotropy of the Fermi velocity:

$$E_{e,h} - \Delta E = \hbar v_{ph}^{(e,h)}(\varphi)|\mathbf{k} - \mathbf{k}_0|, \tag{9}$$

where $\varphi$ is the polar angle of $\mathbf{k} - \mathbf{k}_0$. The electron and hole phase velocities are

$$v_{ph}^{(e,h)}(\varphi) = v_0(\varphi) \pm \delta v(\varphi) \tag{10}$$

with

$$\hbar v_0(\varphi) = \hbar v_F (1 + \bar{\varepsilon}\tilde{d}/2 + \varepsilon_\Delta \tilde{g}_2 \cos 2\varphi/2 + \varepsilon_{xy}\tilde{g}_2 \sin 2\varphi) + E_{u2}u_y \cos 2\varphi + E_{u2}u_x \sin 2\varphi,$$

$$\delta v(\varphi) = 2v_F a_2 (\varepsilon_\Delta \cos\varphi - 2\varepsilon_{xy} \sin\varphi). \tag{11}$$

As we see, the spectrum of strained graphene is essentially anisotropic with anisotropy brought about solely by $H_{\varepsilon k}$ and $H_{uk}$ terms of Hamiltonian. Moreover, acoustic strain breaks equivalence of $\mathbf{k}$ and $-\mathbf{k}$ states and the symmetry of electron and hole spectra.

The availability of the results of first-principles calculations [21] for the case of acoustic strain allows to extract the values of the parameters of $H_{\varepsilon k}$. Specifically, the strain dependences of the Fermi velocity allows us to deduce that $a_2 \approx 0.2$, $\tilde{d}_2 \approx -1.25$, and $\tilde{g}_2 \approx -2.14$. Note, that the spectrum of Eq. (7-11) agrees well with the first-principles calculations for strain as high as 10%. The obtained numerical values of $\tilde{d}_2$, and $\tilde{g}_2$ confirm our rough estimate for the relative contributions of the first and second-order perturbation terms in Eq. (1). To best of our knowledge, there is no first-principles calculations of the graphene spectrum with optical strain, which does not allow us to determine the corresponding constants of $H_u$ and $H_{uk}$.

Let us consider now new effects arising for the case of graphene placed at the substrate. The most important one is opening of the gap in the spectrum equal to $\delta E = 2E_{op}^{(u)} u_z$. It is important to note that this is in agreement with the results of [11] and [42], suggesting that gap in the spectrum appears if the two sublattices of graphene lattice are nonequivalent. It is easy to see that only out of plane optical strain of graphene placed at a substrate breaks this equivalence. The terms with acoustic out of plane displacement give rise to the additional shift of the whole spectrum $\Delta E$ and to the isotropic renormalization of the Fermi velocity:

$$\Delta E = E_{ac}\bar{u}_z, \quad \hbar \delta v_0(\varphi) = E_{a1}\bar{u}_z. \tag{12}$$

Naturally, the values of the constants in Eqs. (4) and (6) depend on the character of interaction of graphene sheet with substrate and it is supposed that they are smaller then the analogous in-plane constants.

4. **Generalization of the Hamiltonian for the case of spatially nonuniform strain**

The developed approach can be generalized for the case of nonuniform strain, which is smooth on the scale of the lattice constant. This is especially important for $H_{\varepsilon k}$ and $H_{uk}$, since $\mathbf{k}$ operator does not commutate with $\varepsilon_{ij}(\mathbf{r})$ and $u_j(\mathbf{r})$. In particular, this situation is inherent for consideration of the electron-long wavelength phonons interaction in graphene. To consider the non-uniform strain, the coordinate transformation method analogous to that used in the case of uniform strain can be applied. It was shown [33] that one should change $\varepsilon_{ij}$ and $u_j$ in $H_\varepsilon$ and $H_u$ to $\varepsilon_{ij}(\mathbf{r})$ and $u_j(\mathbf{r})$ respectively and replace the products $\varepsilon_{ij}k_l$ and $u_j k_i$ in $H_{\varepsilon k}$ and $H_{uk}$ by the corresponding symmetric combinations $\{\varepsilon_{ij}, k_l\} = -i\hbar \left[ \varepsilon_{ij}(\mathbf{r}) \frac{\partial}{\partial r_l} + \frac{1}{2} \frac{\partial \varepsilon_{ij}(\mathbf{r})}{\partial r_l} \right]$ and $\{u_j, k_l\} = -i\hbar \left[ u_j(\mathbf{r}) \frac{\partial}{\partial r_l} + \frac{1}{2} \frac{\partial u_j(\mathbf{r})}{\partial r_l} \right]$ where the standard notation $\{a,b\} = (ab + ba)/2$ is used. It is worth to mention that nonuniform strain fields can mimic the orbital effect of magnetic and electric fields [13-17]. This analogy, however, holds only if $H_\varepsilon$ and $H_u$ are considered. $H_{\varepsilon k}$ and $H_{uk}$ contributions make it not valid strictly, and their role in formation of the graphene spectrum under nonuniform strain needs further analysis.

In conclusion, using symmetry-based approach, we derived the effective Hamiltonian of graphene which takes into account spatially nonuniform acoustic and optical strains. Being an important generalization of the previously developed model for uniform acoustic strain [30] the obtained Hamiltonian can serve as a convenient tool for analysis of various graphene properties, such as spectrum under static uniform or spatially non-uniform strain or characteristics of electron-phonon interaction. Comparing our model and previously published results of first-principles calculations, we extract the numerical values of parameters entering the acoustic part of the Hamiltonian. We show that with this selection of parameters, our Hamiltonian reproduces the first-principles results for strain up to about 10%. We addressed also modifications brought about in graphene placed at a substrate. We have shown that in this case out of plane optical strain lifts equivalence of the graphene sublattices which gives rise to the opening of the gap in the spectrum.

**Acknowledgments**

The author acknowledge the hospitality of the Abdus Salam International Centre for Theoretical Physics (Trieste), where this work was completed. I am also thankful to V A Kochelap and V I Sheka for helpful discussions.

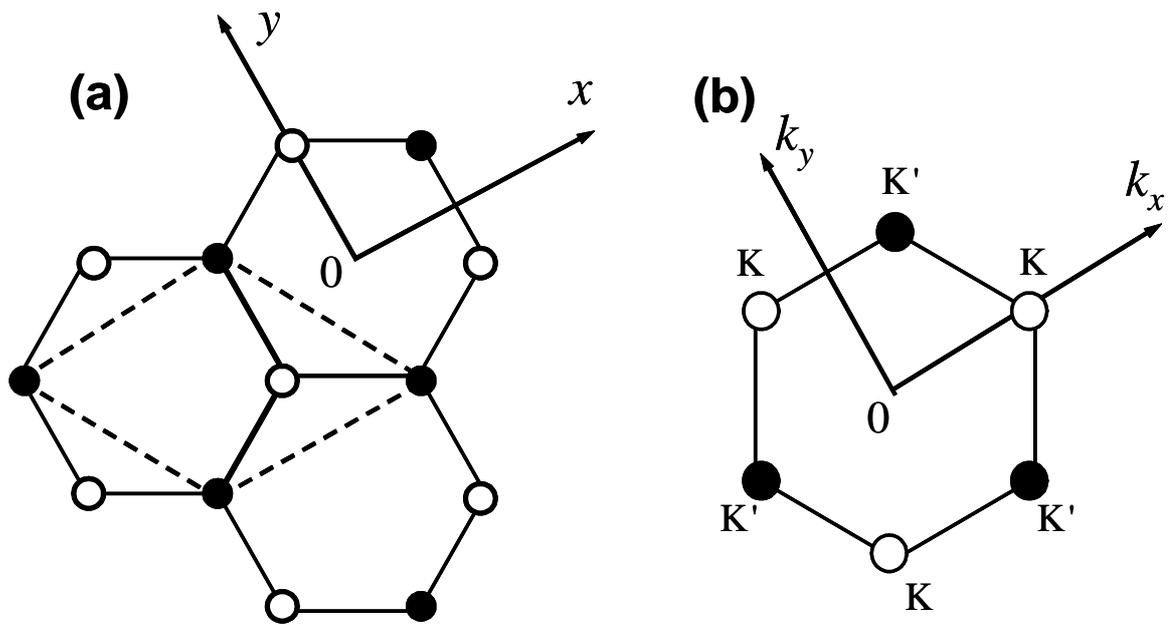

Figure1. (a) The honeycomb lattice of graphene. The carbon sites belonging to two equivalent sublattices are denoted by solid and hollow circles. The dash line marks the primitive cell. (b) The first Brillouin zone of graphene; the Dirac points of the graphene spectrum are at K and K' valleys.